\documentclass[conference]{IEEEtran}
\IEEEoverridecommandlockouts
\usepackage{listings}
\usepackage{cite}
\usepackage{amsmath,amssymb,amsfonts}
\usepackage{graphicx}
\usepackage{xcolor}
\usepackage{subfigure}
\usepackage{booktabs}
\usepackage{multirow}
\usepackage{makecell}
\usepackage{threeparttable}
\usepackage{enumitem} 
\usepackage{url}
\newtheorem{example}{Example}[section]

\def\BibTeX{{\rm B\kern-.05em{\sc i\kern-.025em b}\kern-.08em
		T\kern-.1667em\lower.7ex\hbox{E}\kern-.125emX}}

\begin{document}
	
\title{Testing Quantum Programs with Multiple Subroutines}

\author{
\IEEEauthorblockN{Peixun Long}
	\IEEEauthorblockA{\textit{State Key Laboratory of Computer Science} \\
		\textit{Institute of Software, Chinese Academy of Science} \\
		\textit{University of Chinese Academy of Science}\\
		Beijing, China \\
		longpx@ios.ac.cn}
	\and
	\IEEEauthorblockN{Jianjun Zhao}
	\IEEEauthorblockA{\textit{Kyushu University} \\
		Fukuoka, Japan \\
		zhao@ait.kyushu-u.ac.jp}
}

\maketitle
\thispagestyle{plain}
\pagestyle{plain}

\begin{abstract}
Errors in quantum programs are challenging to track down due to the uncertainty of quantum programs. Testing is, therefore, an indispensable method for assuring the quality of quantum software. Existing testing methods focus only on testing quantum programs with quantum circuits or single subroutines and, therefore, cannot effectively test quantum programs with multi-subroutines. In this paper, we first discuss several critical issues that must be considered when testing multi-subroutine quantum programs and point out the limitations and problems with existing testing methods. We then present a novel framework for testing multi-subroutine quantum programs that allow for both unit and integration testing. Our framework includes three novel test coverage criteria for the equivalent class partition of quantum variables to guide our testing tasks and techniques to test quantum programs with several common patterns. We also discuss how to generate test cases based on our framework. To evaluate the effectiveness of our testing framework, we implemented a tool called QSharpTester for testing Q\# programs with multiple subroutines. We used it to conduct experiments on hundreds of mutation programs deriving from seven original Q\# programs. The experimental results show that our testing methods can deal with broader types of quantum programs than existing ones and perform well on almost all faulty mutation programs.

\end{abstract}


\section{Introduction}
\label{sec:introduction}
Quantum computing is a rapidly developing field expected to make breakthroughs in many areas~\cite{national2019quantum}. It uses the principles of quantum mechanics to process information and computing tasks. Compared with classical algorithms, quantum algorithms can solve some specific problems with acceleration~\cite{deutsch1985quantum,grover1996fast,shor1999polynomial}. 
With the development of quantum hardware devices and algorithms, it becomes increasingly important to develop high-quality quantum software. 

Recently, many quantum programming languages have been proposed, such as Q\#~\cite{svore2018q}, Qiskit~\cite{gadi_aleksandrowicz_2019_2562111}, Scaffold~\cite{abhari2012scaffold}, Quipper~\cite{green2013introduction}, and Sliq~\cite{bichsel2020silq}, for supporting writing programs that run on quantum simulators or real quantum devices. 
However, due to the specific features of quantum programs, such as superposition, entanglement, and non-cloning theorem, errors in quantum programs are challenging to track down~\cite{miranskyy2020your,huang2019statistical}.
%
Although several testing methods have been proposed for quantum programs~\cite{ali2021assessing,honarvar2020property,wang2018quanfuzz,li2020projection,miranskyy2019testing,abreu2022metamorphic,wang2021generating,wang2021application}, they still have some limitations: (1) Existing methods migrate techniques for classical programs into quantum programs and lack the consideration of quantum nature. (2) Existing methods consider mostly single quantum programs, but multi-subroutine quantum programs are more common in practice and lack systematic support for testing them. (3) The definitions of input, output, and program specifications in existing testing methods limit the scope of their application to test multi-subroutine quantum programs.

In this paper, we first discuss several critical issues that must be considered when testing multi-subroutine quantum programs and point out the limitations and problems with existing testing methods.
Then, we propose a systematic test process and a framework to support the testing tasks for multi-subroutine quantum programs according to these properties. We discuss IO analysis, equivalence class partition, coverage criteria, and test case generation in succession. In IO analysis, we introduce the \textit{IO mark} to denote the input and output of each subroutine. In equivalence class partition and coverage criteria, we propose three novel coverage criteria for quantum variables and discuss the combinations of variables. In test case generation, we discuss the principle of choosing test inputs and how to create the required input and detect output. We also discuss how to test quantum programs with two common patterns. To support our testing methods and future testing tasks, we implement QSharpTester, an efficient testing tool for testing Q\# programs with multi-subroutines.

To evaluate the effectiveness of our testing methods, we use a set of Q\# benchmark programs, including seven original correct programs and their 244 faulty mutations with four mutation types. We use Q\# and its simulator because Q\# is a program-level rather than circuit-level programming language~\cite{svore2018q} to support complex quantum programs. The results of benchmark programs show that our methods can find almost all faulty mutations and some equivalent mutations. The results also show that our methods can finish testing tasks in an acceptable time on a personal computer with a Q\# simulator. We also compare our testing methods with existing methods. Our methods have advantages in (1) supporting the testing of broader types of quantum programs, (2) finding bugs related to quantum natures, and (3) scalability and compatibility.

Our paper makes the following contributions:

\begin{itemize}[leftmargin=2em]
    \item \textbf{Critical issues in testing quantum programs:} We discuss and clarify some critical issues in testing muti-subroutine quantum programs, which have been either neglected or rarely discussed by previous research.
    
    \item \textbf{Testing coverage criteria:} We present three novel testing coverage criteria, namely, \textit{classical-superposition partition}, \textit{classical-superposition-mixed partition}, and \textit{entanglement partition} for equivalence class partitioning of quantum programs.
    
    \item \textbf{Testing framework:} We present a framework for testing multi-subroutine quantum programs. Based on the framework, we also implement QSharpTester, a tool for testing Q\# programs with multi-subroutines.
    
    \item \textbf{Evaluation:} We use a set of benchmark programs to evaluate the effectiveness of our testing methods. The experimental results show that our testing methods can deal with broader types of quantum programs than existing ones and perform well on almost all faulty mutation programs.
    

\end{itemize}

\vspace{2mm}
The rest of the paper is organized as follows. Section~\ref{sec:background} introduces some basic concepts of quantum computation. We discuss how the properties of multi-subroutine quantum programs influence the testing tasks in Section~\ref{sec:basic-concepts}. Section~\ref{sec:testing-framework} presents the details of our testing methods, and Section~\ref{sec:QSharpTester} discusses the tool support for testing multi-subroutine quantum programs with QSharpTester. An experimental evaluation is given in Section~\ref{sec:evaluation}. We discuss related works in Section~\ref{sec:related-work}, and the conclusion is given in Section~\ref{sec:conclusion}.

\section{Background}
\label{sec:background}
We first introduce basic information on quantum computation, quantum algorithms, and quantum programming languages.

\subsection{Basic Concepts of Quantum Computation}
\label{subsec:basic-concepts}
A quantum bit, or \textit{qubit} for short, is the basis of quantum computation. We know that a classical bit has two values: 0 and 1. A qubit also has two \textit{states} with the form $\left|0\right>$ and $\left|1\right>$ like the classical bit, and it can contain a \textit{superposition} between \textit{basis states}. The general state of a qubit is $a\left|0\right>+b\left|1\right>$, where $a$ and $b$ are two complex numbers called \textit{amplitudes} that satisfy $|a|^2+|b|^2=1$. Amplitudes represent the probability proportion of each basis state. For more than one qubit, the basis states are similar to binary strings. For example, a two-qubit system has four basis states: $\left|00\right>$, $\left|01\right>$, $\left|10\right>$ and $\left|11\right>$, and the general state is $a_{00}\left|00\right>+a_{01}\left|01\right>+a_{10}\left|10\right>+a_{11}\left|11\right>$, where $|a_{00}|^2+|a_{01}|^2+|a_{10}|^2+|a_{11}|^2=1$. The state can also be written as a column vector $[a_{00},a_{01},a_{10},a_{11}]^T$. The above states are all \textit{pure states}. Sometimes a quantum system may have a distribution over several pure states rather than a certain state, called a \textit{mixed state}.

In quantum devices, the information of qubits can only be obtained by \textit{measurement}. Measuring a quantum system will get a classical value with the probability of corresponding amplitude. Then the state of the quantum system will collapse into a basis state according to obtained value. For example, measuring a qubit $a\left|0\right>+b\left|1\right>$ will get result 0 and collapse into state $\left|0\right>$ with probability $|a|^2$ and get result 1 and collapse into state $\left|1\right>$ with probability $|b|^2$. This property brings uncertainty and influences testing work for quantum programs. Another important property is \textit{entanglement}. When two qubits are entangled, the entire state cannot be departed as two irrelevant qubits. The Bell state $\left|\beta_{00}\right>=\frac{1}{\sqrt 2}(\left|00\right>+\left|11\right>)$ is a typical entangled state.

Quantum computing is performed by applying proper quantum gates on qubits. An $n$-qubit quantum gate can be represented by a $2^n \times 2^n$ unitary matrix $G$. Applying gate $G$ on quantum state $\left|\psi\right>$ will obtain state $G\left|\psi\right>$. There are several basic quantum gates: single-qubit gates $X$, $Z$, $H$, and $R_1(\theta)$, and two-qubit gate CNOT. The matrices of these gates are shown in the following:

\small
\begin{equation}
\notag
X=\left[\begin{array}{cc}
0 & 1\\
1 & 0
\end{array}
\right], \hspace*{2mm}
Z=\left[\begin{array}{cc}
1 & 0\\
0 & -1
\end{array}
\right], \hspace*{2mm}
H=\frac{1}{\sqrt{2}} \left[\begin{array}{cc}
1 & 1\\
1 & -1
\end{array}
\right]
\end{equation}

\vspace{-3mm}

\begin{equation}
\notag
R_1(\theta)=\left[\begin{array}{cc}
1 & 0\\
0 & e^{i\pi \theta}
\end{array}
\right], \hspace*{2mm}
\mathrm{CNOT}=\left[\begin{array}{cccc}
1&0&0&0\\
0&1&0&0\\
0&0&0&1\\
0&0&1&0
\end{array}
\right]
\end{equation}

\begin{figure}[h]
	\centering
	\includegraphics[scale=0.6]{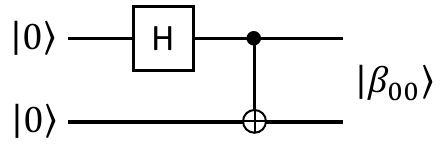}
	\caption{A quantum circuit for preparing the Bell state, containing an $H$ gate and a CNOT gate.}
	\label{FIGbell}
\end{figure}

\normalsize
A quantum circuit is a popular model to express the process of quantum computing. Every line represents a qubit in a quantum circuit model, and a sequence of operations is applied from left to right. For example, Figure~\ref{FIGbell} shows a quantum circuit for preparing the Bell state $\left|\beta_{00}\right> = \frac{1}{\sqrt 2}(\left|00\right>+\left|11\right>)$, which contains an $H$ gate and a CNOT gate.

\subsection{Quantum Algorithms}
Quantum algorithms use quantum mechanics to solve computational problems. This section introduces some quantum algorithms, which will be discussed in the following sections.

A fundamental quantum algorithm is Grover Search (GS)~\cite{grover1996fast}, which has also been proven to be faster than its classical counterparts. GS only needs $O(\sqrt{n})$ times of oracle queries for searching a database with $n$ elements~\cite{10.5555/870802}.

Quantum Fourier Transform (QFT) is another fundamental quantum algorithm that performs the following transformation:

\begin{equation}
    \notag
    \left|j\right> \rightarrow \frac{1}{\sqrt{2^n}}\sum_{k=0}^{2^n-1}e^{2\pi i j k / 2^n}\left|k\right>
\end{equation}

\noindent The Fourier coefficients are encoded in the amplitudes. QFT requires only $O(n^2)$ steps for $n$-qubits input, whereas the classical fast Fourier transform (FFT) still needs $O(n2^n)$ steps. QFT has an inverse transform, which can be denoted as IQFT.

\begin{figure}[h]
	\centering
	\includegraphics[scale=0.6]{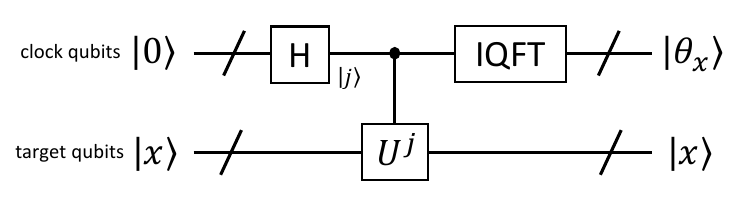}
	\caption{The quantum circuit for QPE algorithm.}
	\label{FIGqpe}
\end{figure}

Quantum Phase Estimation (QPE) is a typical application of QFT to attain the eigenvalue of a given eigenvector of a target unitary operation. Let $U$ be the unitary operation, and it has an eigenvector $\left|x\right>$ with eigenvalue $e^{2\pi i \theta_x}$ (The length of the eigenvalue of a unitary is 1). QPE estimates $\theta_x$ and stores its binary representation in multi-qubit quantum state $\left|\theta_x\right>$. Figure~\ref{FIGqpe} gives the overall circuits of the QPE algorithm, which requires two groups of qubits: \textit{target qubits} and \textit{clock qubits}. The target qubits are initialized in state $\left|x\right>$, and the clock qubits are initialized in an \textit{all-zero} state. The binary representation of $\theta_x$ will be output in the clock qubits.

Quantum Order Finding (QOF) is a typical application of QFT and QPE. Shor's Quantum Factoring (QF) and discrete logarithms~\cite{shor1999polynomial} are based on QOF. QF is a vital algorithm because it can break some cryptosystems based on the complexity of large number decomposition, such as RSA.

\subsection{Quantum Programming Languages}
Several quantum programming languages are available recently, such as ProjectQ~\cite{haner2016high}, Scaffold~\cite{abhari2012scaffold,javadiabhari2014scaffcc}, Q{\sc wire}~\cite{ paykin2017qwire}, Qiskit~\cite{gadi_aleksandrowicz_2019_2562111}, and Q\#~\cite{svore2018q}. Some of these languages, such as the Python-based ProjectQ, are extensions of classical programming languages, and some, such as Q\#, are independent of their classical counterparts. Some languages, such as Q{\sc wire}, are lower level and suitable for describing quantum circuits. In contrast, others, such as Q\#, are higher level and capable of implementing complex function calls. In addition, almost all quantum programming languages provide simulators that allow programmers to run quantum programs on a classical computer. A few programming languages even provide interfaces to real quantum hardware. For example, Qiskit has an interface to IBM's quantum hardware.

\section{The properties of multi-subroutine quantum programs}
\label{sec:basic-concepts}

We next discuss some critical properties of multi-subroutine quantum programs and try to answer the following research question.

\begin{itemize}
    \item How do the properties of multi-subroutine quantum programs influence testing tasks?
\end{itemize}

\subsection{Quantum Circuits and Quantum Programs}
\label{subsec:CircuitProgram}

Conventionally, a quantum circuit is used to describe a quantum algorithm, as shown in Figure~\ref{FIGqpe} for the QPE algorithm. However, quantum circuits are not always equivalent to quantum programs, and the quantum circuit model has some limitations in describing general quantum programs. A typical limitation is quantum-classical-hybrid programming. The quantum circuit model is hard to describe quantum programs with classical variables or control flows (such as \textit{if} statements or \textit{while} loop statements).

Many practical quantum algorithms contain both classical and quantum parts. A case is where the input contains both classical and quantum variables. A typical example is Parameterized Quantum Circuit (PQC)~\cite{marcello2019PQC}, widely used in quantum machine learning. Another case is that the program contains both classical and quantum subroutines. For example, the unique quantum subroutine of the QF program is QOF, and other subroutines are classical. So the description of QF is, to some extent, like a classical algorithm rather than a quantum circuit.

\begin{figure}
	\centering
	\subfigure[$n=1$]{\includegraphics[scale=0.25]{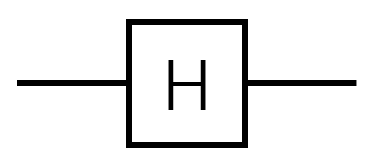}}
	\subfigure[$n=2$]{\includegraphics[scale=0.5]{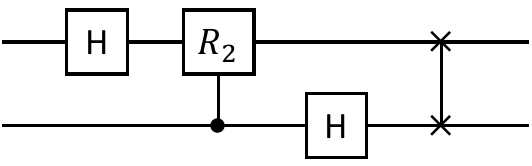}}
	\\
	\centering
	\subfigure[$n=3$]{\includegraphics[scale=0.5]{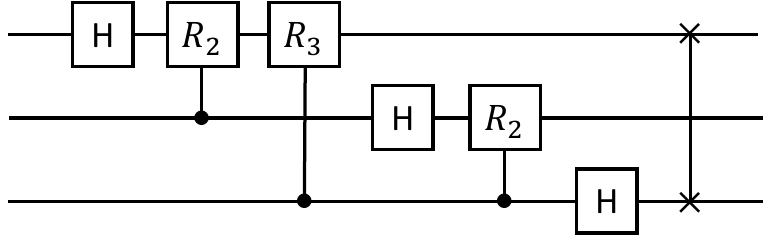}}
	
	\caption{Corresponding circuits of QFT program with $n=1,2$ and $3$, where $n$ is the number of qubits.}
	\label{FIGqft}
\end{figure}

A general quantum program is usually not on a fixed scale, which often corresponds to a family of quantum circuits rather than a specific circuit. Although we can represent it as an unfixed circuit, as Figure~\ref{FIGqpe} shows, it is a high-level representation, omitting many details. As shown in Figure~\ref{FIGqft}, the QFT program corresponds to different circuits when the number $n$ of qubits is different, similar to the classical counterpart of the quantum Boolean circuit~\cite{arora2009computational}, an essential computational model in computational complexity. A classical algorithm usually corresponds to \textit{a uniform family of Boolean circuits} rather than just a circuit.


This paper focuses on testing quantum programs rather than quantum circuits and considering both quantum and classical code in a quantum program. Moreover, the differences between quantum programs and quantum circuits motivate us to reconsider the current quantum program testing process.

%



\subsection{Subroutines}
\label{subsec:Subroutines}

A complete quantum program is usually composed of a number of subroutines. Listing~\ref{LISTqft} shows an implementation of the QFT program in Q\#. In the program, \texttt{CRk} and \texttt{Reverse} are two subroutines called by the upper-level subroutine \texttt{QFT}. Besides this direct call, another way to organize subroutines is to take the subroutine as an input parameter. An example is Grover Search~\cite{grover1996fast}, which needs a subroutine to identify a solution and uses the subroutine as a black box. We call the subroutine an \textit{oracle}. Listing~\ref{LISTqpe} shows another example \texttt{QPE}, which requires the target quantum operation as an input parameter (\texttt{Upower} in lines 1 and 8).


\begin{figure}
\scriptsize{
\begin{lstlisting}[
	xleftmargin=4mm,
	%xrightmargin=2mm,
	caption={The Q\# code of QFT program},
	label={LISTqft}
	]
operation CRk(qctrl: Qubit, k: Int, qtarget: Qubit) 
          : Unit is Adj {
    body(...) {
        let theta = 2.0*Math.PI()/Convert.IntAsDouble(2^k);
        Controlled R1([qctrl], (theta, qtarget));
    }
    adjoint auto;
}

operation Reverse(qs : Qubit[]) : Unit is Adj {
    body(...) {
        let n = Length(qs);
        for i in 0 .. n / 2 - 1
        { SWAP(qs[i], qs[n - i - 1]); }
    }
    adjoint auto;
}

operation QFT(qs : Qubit[]) : Unit is Adj {
    body(...) {
        let n = Length(qs);
        for i in 0 .. n - 2 {
            H(qs[i]);
            for j in i + 1 .. n - 1 {
                let k = j - i + 1;
                CRk(qs[j], k, qs[i]);   // Call "CRk"
            }
        }
        H(qs[n - 1]);
        Reverse(qs);    // Call "Reverse"
    }
    adjoint auto;
}
\end{lstlisting}
}
\end{figure}

\begin{figure}
\scriptsize{
\begin{lstlisting}[
	xleftmargin=4mm,
	%xrightmargin=2mm,
	caption={The Q\# code of QPE program},
	label={LISTqpe}
	]
operation QPE(Upower : (Int,Qubit[])=>Unit is Adj+Ctl,
           qsclock : Qubit[], qstarget : Qubit[])
           : Unit is Adj {
    body (...) {     //Original operation
        let n = Length(qsclock);
        MultiH(qsclock);
        for i in 0 .. n - 1 {
            Controlled Upower([qsclock[n - i - 1]],
                              (2 ^ i, qstarget)); 
        }
        Adjoint QFT(qsclock);
    }
    adjoint auto;  //Generate adjoint version of QPE by Q#.
}
\end{lstlisting}
}
\end{figure}

Some variants deriving from the original subroutine are needed in many quantum programs. For example, as shown in Listing~\ref{LISTqpe}, the \texttt{QPE} program calls~\texttt{Adjoint QFT} (line 11), which is the inverse of \texttt{QFT}. It also calls \texttt{Controlled Upower} (lines 8-9), which is the controlled version of \texttt{Upower}. Generally, there are three important variants of a subroutine in quantum programs: \textit{inverse}, \textit{controlled}, and \textit{power}, and these variants can also be combined. Formally, for a unitary operation $U$, the \textit{inverse} variant is $U^{-1}$, the \textit{controlled} is $CU$, and the \textit{power} is $U^n$.

Some high-level quantum programming languages, such as Q\#~\cite{svore2018q} and isQ~\cite{Guo2022isQ}, support the generation and management of these three variants. As an example, Listing~\ref{LISThp} gives an implementation of $H^n$, the power of the $H$ gate, using these three variants. The power $n$ is implemented as an extra \texttt{Int} parameter \texttt{power}. The result from $H^2=I$, $H^n$ equals $H$ if $n$ is odd; otherwise, it equals $I$ (line 5). The inverse, controlled, and inverse-and-controlled variants can be generated automatically by Q\# (lines 7-9, and Q\# uses the keyword ``\texttt{adjoint}" to represent inverse operation). Because of the wide use of these three variants, it is necessary to propose specific techniques to test them. We will consider this issue in Section~\ref{subsec:testvariants}.

\begin{figure}
\scriptsize{
\begin{lstlisting}[
	xleftmargin=4mm,
	%xrightmargin=2mm,
	caption={The implementation of the power variant of the H gate. The adjoint, controlled, and adjoint controlled versions are generated by Q\#.},
	label={LISThp}
	]
operation Hpower(power : Int, q : Qubit)
                : Unit is Adj + Ctl {
    body (...) {
        //If pow is odd, apply H, otherwise I
        if (power % 2 == 1) { H(q); }
    }
    adjoint auto;
    controlled auto;
    adjoint controlled auto;
}
\end{lstlisting}
}
\end{figure}


\subsection{Input and Output}
\label{subsec:IO}

Previous research~\cite{ali2021assessing, wang2021generating, wang2021application} defines the input and output of a quantum program as a subset of all used qubits. Such a definition has some problems when testing multi-subroutine quantum programs. First, it ignores the classical variables, an essential part of many quantum programs. Second, it regards the total input/output as an entity, ignoring the detailed logical meaning.
This paper considers both the quantum and classical input and output of a quantum program. It identifies the input and output by logic meaning, where the quantum variables can be seen as a special type. For example, consider a quantum adder program $\left|x\right>\left|y\right>\rightarrow\left|x\right>\left|x+y\right>$. The quantum registers of $\left|x\right>$ and $\left|y\right>$ are logically two sets of qubits. Therefore, the input should contain two quantum variables $\left|x\right>$ and $\left|y\right>$, rather than taking all qubits as one input variable.
For a target program under test, correctly identifying the inputs and outputs of all program subroutines is one of the prerequisites for a successful testing task. We will give a more detailed discussion in Section~\ref{subsec:IOanalysis}.

\subsection{Program Specification}
\label{subsec:PS}

Program specification gives the expected behavior of the program and is the basis for testing execution. The purpose of testing is to check whether a program converts the given input into a specific output according to the program specification. Previous research~\cite{wang2021generating, ali2021assessing, wang2021application} defined the program specification as the expected probability distribution of the output values under given input values. However, this definition implies that we must have a measurement at the end of each subroutine to convert quantum states into probability distributions. This restriction is not suitable for testing a multi-subroutine program. In a multi-subroutine program, a subroutine may only transform a quantum state into another state, which is not necessary to measure the state at the end of the subroutine.

On the other hand, it is better to consider the specification of quantum states rather than a probability distribution. Generally, for some given inputs, the program specification gives the expected output for each input. This relationship can be represented as a formula in most cases, and we call it \textit{formula-based program specification}. For example, QFT has the following formula description:

\begin{equation} \label{EQUqftform}
	\left|j\right> \rightarrow \frac{1}{\sqrt{2^n}}\sum_{k=0}^{2^n-1}e^{2\pi i j k / 2^n}\left|k\right>
\end{equation}


\noindent where $\left|j\right>$ is a classical state with integer $j$. Formula (\ref{EQUqftform}) only gives the transformation under classical state inputs. Fortunately, QFT is a unitary transform, so the expected output under general input can be deduced by linearity.




\subsection{Quantum State Generation and Detection}
\label{subsec:GenerationDetection}

Testing quantum programs relies on the processing of quantum variables. However, because of the properties of quantum states, it is challenging to generate and detect quantum states compared to the generation and detection of classical states.

In the classical case, putting some parameters into a computer is straightforward. It is easy to convert a natural language description of the input (such as a block of text or a decimal number) into a binary string that a computer can recognize according to some predetermined coding rules. The input process is to convert the binary string into the state of internal components of a computer (e.g., high and low voltages represent 1 and 0, respectively). However, quantum variables not only have classical states like binary strings but also have superposition states or mixed states. Some states which are easily described in natural language may be difficult to prepare on quantum devices. Furthermore, it has been proved that some states need high costs to prepare~\cite{knill1995approximation}.

After running the program under a given input, we must compare the actual output with the expected output. Reading out the contents of classical memories is also straightforward. However, we cannot avoid measurement to obtain information on quantum variables, which may change the state of quantum variables and disrupt the execution of programs.


\section{Testing Framework}
\label{sec:testing-framework}

We next present a systematic test process and a testing framework for testing multi-subroutine quantum programs.

\subsection{The Overall Test Process}
\label{subsec:OverallProcess}

\begin{figure}[t]
	\centering
	\includegraphics[scale=0.75]{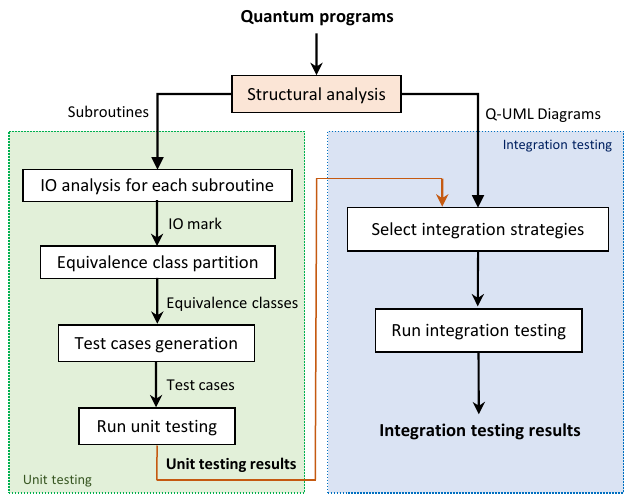}
	\caption{The overall process of testing quantum programs with multiple subroutines.}
	\label{FIGoverall}
\end{figure}

Figure~\ref{FIGoverall} shows the overall process of testing quantum programs with multiple subroutines. In this process, given a quantum program as input, the first step is to analyze the structure of the whole program and find its subroutines and their organization. To this end, we use Q-UML~\cite{Perez-Delgado2020quantum} to specify the structure of the program. After that, unit testing is performed on each subroutine, and then integration testing on the whole program. In unit testing, we first analyze the inputs and outputs of each subroutine to obtain the necessary information for unit testing. We present a novel method called \textit{IO marks} to represent this information. We then partition the input space of each subroutine into several equivalence classes and generate test cases according to these equivalence classes for the subroutine. We introduce three novel coverage criteria. After that, we run the unit test to obtain the testing result for each subroutine. The next step is to perform integration testing by selecting proper integration methods to run and obtain the testing results for the whole program.

In the rest of this section, we will discuss each part of our test process, present tools and methods to deal with them, and provide examples to show how to use our testing methods.

\subsection{IO Analysis}
\label{subsec:IOanalysis}

As discussed in Section~\ref{subsec:IO}, analyzing the input and output is the origin of testing each subroutine. For a concrete subroutine, only the variables that users should assign are regarded as input. Similarly, only the variables users are interested in after running are regarded as output. For example, in many quantum algorithms, some qubits should always be initialized with an \textit{all-zero state} $\left|0\ldots0\right>$. These qubits should not be regarded as input. Another typical case is a qubit array, such as \texttt{qs} in Listing~\ref{LISTqft} (line 19). Like an array of integers, a qubit array contains both length and data. Therefore, when identifying the input and output of a subroutine, a qubit array should be regarded as two variables: length and quantum state.


To represent the input and output of a quantum program, we introduce the \textit{IO mark}, which can help developers design test cases. The general form of the IO mark is:

\vspace{-1mm}
\begin{center}
\textit{program} : (\textit{input variables}) $\rightarrow$ (\textit{output variables'})
\end{center}
\vspace{-1mm}

We use \underline{underline} to denote the input parameters of subroutine-type and \textbf{bold font} to denote quantum variables. 
We add an ``apostrophe" (') on each output variable to distinguish input and output. If a variable~\texttt{var} is both input and output, we denote the input and output as the same name, i.e., \textit{var} is the input, and \textit{var'} is the output. Example~\ref{example:io-mark} shows how to perform IO analysis and use IO marks.

\begin{figure}
	\centering
	\includegraphics[scale=0.6]{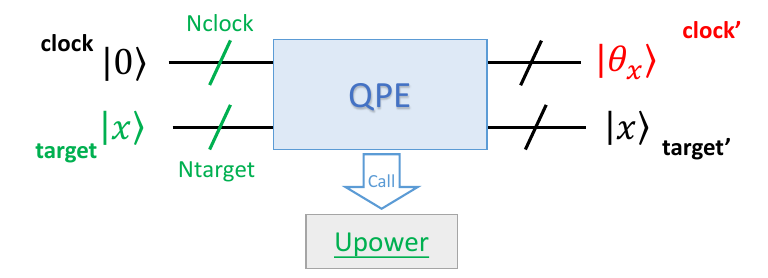}
	\caption{The diagram represents the input and output of \texttt{QPE} program, where the input variables are denoted in green, and the output variable is denoted in red.}
	\label{FIGqpeio}
\end{figure}

\vspace{2mm}
\begin{example}\label{example:io-mark}
\noindent
IO analysis for \texttt{QPE} program.

The \texttt{QPE} program in Listing~\ref{LISTqpe} has three parameters: \texttt{Upower}, \texttt{qsclock}, and \texttt{qstarget}, while the latter two are qubit-array types. On the input side, \texttt{qsclock} contains two variables: the length ``Nclock" and the data ``\textbf{clock}." It is similar to \texttt{qstarget}, which contains two variables as well: the length ``Ntarget" and the data ``\textbf{target}." On the output side, they should be denoted as ``\textbf{clock'}" and ``\textbf{target'.}" \texttt{Upower} is a parameter of subroutine type and should be denoted as ``\underline{\texttt{Upower}}." Figure~\ref{FIGqpeio} shows the structure, input, and output of the \texttt{QPE} program.

In the program, the quantum state \textbf{clock} is always initialized with an all-zero state, so it is not an input. In most applications of the \texttt{QPE} program, we do not care about the post-state \textbf{target'}, so it is not output. In Figure~\ref{FIGqpeio}, the input variables are shown in green, while the output variable is red. The IO mark can be written as:

\vspace{2mm}
\begin{center}
	\texttt{QPE} : (Nclock, Ntarget, \underline{\texttt{Upower}}, \textbf{target}) $\rightarrow$ (\textbf{clock'})
\end{center}

\end{example}

\subsection{Equivalence Class Partition and Testing Coverage Criteria}
\label{subsec:CoverageCriteria}

A coverage criterion is a set of rules used to help determine whether a program is adequately tested by a test suite and guides the test process~\cite{ammann2016introduction}. Some coverage criteria for quantum programs have been proposed, such as Quito (Quantum Input-output coverage)~\cite{ali2021assessing}, to provide an objective measure of the test quality of the programs. However, these coverage criteria are based only on the specific input, output, and program specification definitions. As discussed in Section~\ref{subsec:IO} and Section~\ref{subsec:PS}, these criteria have some limitations when applied for testing multi-subroutine quantum programs. For example, the size of the generated test suite from Input Coverage (IC) in~\cite{ali2021assessing} is $|{VD}_I|$, where ${VD}_I$ is the subset of valid input values. In most cases, $|{VD}_I| = O(2^n)$, where $n$ is the number of input qubits, which means that the IC criterion becomes unavailable when testing large-scale quantum programs. The Output Coverage (OC) and Input-Output Coverage (IOC) also have a similar problem of exponential complexity. As a result, these coverage criteria are unsuitable for guiding the overall test process for testing complex quantum programs with multiple subroutines. In this paper, we define a novel class of coverage criteria for partitioning quantum inputs of quantum programs.


Equivalence class partitioning is an important testing method~\cite{myers1979art}. The basic idea is to partition input space into several ``equivalence classes" by logical meaning, and all cases in one class have the same effect in uncovering bugs. Such an idea can be applied to quantum program testing as well. The partitioning of quantum state input often depends on the specific program, but some general principles are still available.

We define a quantum variable as a subset of all qubits in a logical meaning and the ``data" of this variable as the quantum state on these qubits. Generally, there are three types of quantum states: \textit{classical state}, \textit{superposition state}, and \textit{mixed state} (see Section~\ref{subsec:basic-concepts}). So a natural partition criterion is to divide a quantum variable by state types, and we call it ``classical-superposition-mixed partition (CSMP)". However, many quantum programs will not run under input with mixed states. For these programs, we can omit the coverage of mixed states, leading to a ``classical-superposition partition (CSP)". We define two novel coverage criteria as follows.

\vspace*{2mm}
\noindent\textbf{Criterion 1} \textit{classical-superposition partition} (CSP) For each input variable of each quantum state type, partition it into classical state input and superposition state input.

\vspace*{1mm}
\noindent\textbf{Criterion 2} \textit{classical-superposition-mixed partition} (CSMP) For each input variable of each quantum state type, partition it into classical state input, superposition state input, and mixed state input.

\vspace*{2mm}

During the testing, which criterion should be used is determined by the usage of the target program. If we run a target program under mixed-state input, we can use CSMP. Otherwise, CSP is enough. Note that both classical and superposition states should be covered. There are two reasons why both classical and superposition states should be covered in any case. First, the program specification in formula form gives the expected output states of the program under a specific input state, and the input states in the formula are usually classical (such as formula (\ref{EQUqftform}) in Section~\ref{subsec:PS}), so covering classical states is to check the specification directly. Second, superposition is the essential difference between classical and quantum variables, and testing classical input alone cannot ensure the program behavior in superposition input. We will discuss the necessity of covering superposition states in Section~\ref{subsec:necessity}.

The CSP and CSMP criteria are coarse-grained. We often need to partition a concrete quantum program more elaborately according to its logical meaning. Each equivalence class has many input choices, and a good input choice should reveal more potential bugs. For equivalence classes of classical states, we can treat them as classical integers and can migrate classical methods, such as dividing by a boundary value, into the partition. For the equivalence class of superposition states, entanglement is the essential difference between classical and superposition states. Since bugs can occur at any qubit, it is necessary to have entanglement cover every qubit. To this end, we introduce another novel coverage criterion as follows.

\vspace*{2mm}
\noindent\textbf{Criterion 3} \textit{Entanglement Coverage} (EntC). For a set of inputs with superposition states, entanglement must be generated for every qubit.
\vspace*{2mm}


A typical superposition state that satisfies the EntC criterion is of the form $\frac{1}{\sqrt{2}}(\left|x\right>+\left|\bar{x}\right>)$, where $x$ is a binary representation of an integer and $\bar{x}$ is the \textit{bitwise-negation} of $x$. We call such states as \textit{complementary superposition states}.




On the other hand, many quantum programs have more than one input variable, so the combination of variables should also be considered. Several combination coverage criteria between variables have been proposed for classical software testing, such as All Combination Coverage (ACoC)~\cite{cohen1997aetg}, Each Choice Coverage (ECC)~\cite{ammann1994using}, Pair-wise Coverage (PWC)~\cite{burroughs1994improved}, and Base Choice Coverage (BCC)~\cite{ammann1994using,cohen1994automatic}. These coverage criteria can be migrated into quantum programs.

\vspace{2mm}
\begin{example}\label{example:qftpart}
Partitioning and combination for \texttt{QFT} program. The IO mark of the \texttt{QFT} program is:

\begin{center}
    \texttt{QFT} : ($n$, \textbf{qs}) $\rightarrow$ (\textbf{qs'})
\end{center}

As is shown in Listing~\ref{LISTqft}, there are two loops in the \texttt{QFT} program, where the maximum index of the outer loop (line 22) is $n-2$, and the maximum index of the inner loop (line 24) is $n-1$. It is easy to see that 1 and 2 are two boundary values of $n$, so partition $n$ into three parts: $n=1$, $n=2$, and $n \geq 3$. We use CSP to partition \textbf{qs} into a classical state (C) and a superposition state (S). To combine $n$ and \textbf{qs}, we use ACoC on these two parameters. Finally, there are six equivalence classes:

\vspace{1mm}
\begin{center}
\small{
	($n=1$, C)\qquad($n=1$, S)\qquad($n=2$, C)
	
	($n=2$, S)\qquad($n \geq 3$, C)\qquad($n \geq 3$, S)
}
\end{center}
\end{example}

\subsection{Test Case Generation}
\label{subsec:CaseGen}

Having equivalence class partition, we can generate concrete test cases. In classical testing, a test case is defined as the combination of inputs and expected outputs~\cite{ammann2016introduction} as shown below:

\vspace{-1mm}
\begin{center}
test case: (\textit{inputs}, \textit{expected outputs})
\end{center}
\vspace{-1mm}

We choose some pairs of values to fill \textit{inputs} and \textit{expected outputs}. However, for testing quantum programs, we should consider more about the feasibility and how to execute the test cases due to the specific properties of quantum states (see Section~\ref{subsec:GenerationDetection}). Therefore, for each equivalence class of a quantum variable, we choose quantum input states which should satisfy the following three properties simultaneously:

\begin{itemize}
    \item Input state is typical to represent its type;
    \item Input state is easy to prepare;
    \item Corresponding output state is also easy to check.
\end{itemize}

We discuss how to create inputs and detect outputs for quantum variables in the following.

\vspace{2mm}
\subsubsection{Creating quantum inputs}
For general superposition states, if we know the probability distribution $\{p_i\}$ of each amplitude, we can use the algorithm~\cite{Lov2002Creating} to generate the superposition state $\sum_i \sqrt{p_i}\left|i\right>$. However, in most cases, testing general input is not necessary or infeasible because (1) generating general superposition state $\sum_i \sqrt{p_i}\left|i\right>$ is cost; (2) the corresponding output of general input may be difficult to detect. According to our practice, two typical types of superposition states as follows, are helpful in finding bugs and they can be prepared easily:

\begin{itemize}
    \item \textit{uniform superposition} $\frac{1}{\sqrt{2^n}}\sum_{k=0}^{2^n-1}\left|k\right>$;
    
    \item \textit{two-values superposition} $\frac{1}{\sqrt 2}(\left|x\right>+e^{i\theta}\left|y\right>)$, where two classical integers $x,y$ and phase difference $\theta$ can be chosen by users.
\end{itemize}

If we know the ensemble composition, the preparation is simple for mixed-state input. If we only know the density matrix $\rho$, we can use spectrum decomposition to find the ensemble composition~\cite{nielsen2002quantum}.

\vspace{2mm}
\subsubsection{Detecting quantum outputs}
Although, as discussed in Section~\ref{subsec:GenerationDetection}, obtaining information from quantum memories is difficult, there are still some methods and strategies to finish this task, which can be divided into \textit{statistic-based detection} (SBD) methods and \textit{quantum runtime assertions} (QRAs) methods.

SBD means running the target program many times, collecting the measurement information, and then reconstructing the output state from this information. There have been many methods and algorithms in quantum state detection and discrimination~\cite{Stephen2008Discrimination}, which can be used for checking the output states during testing. For example, \textit{state tomography}, \textit{swap test}~\cite{buhrman2001quantum,ekert2002direct}, and \textit{hypothesis test}~\cite{huang2019statistical}. In essence, these methods are based on the repeated running of programs, which require running many times to achieve high precision, so they are costly.

The QRA methods, such as those proposed in ~\cite{li2020projection,liu2020quantum,DBLP:conf/hpca/LiuZ21}, are based on the fact that if the target output state is a member of the measurement basis, the measurement will not destroy the state. By choosing a proper measurement basis, even only once running is enough to get the necessary information. However, according to our practical experience, the QRAs are not always easy to implement for general quantum programs. In many cases, however, the complexity of implementing a quantum runtime assertion is the same as that of implementing the target program, which means that the burden of testers is heavy, and this testing method is also error-prone. Therefore, we still need to use SBD or hybrid methods for programs that cannot construct the runtime assertions easily.

A typical application of QRAs in testing tasks is \textit{transform-based detection} (TBD). Suppose expected output $\left|\psi_{eo}\right>$ can be obtained by applying a simple unitary operation on an all-zero state, i.e., there exists a unitary operation $U_{eo}$ such that $\left|\psi_{eo}\right> = U_{eo}\left|0\ldots0\right>$. If the practical output is $\left|\psi_{eo}\right>$, applying $U_{eo}^{-1}$ on it will obtain an all-zero state, otherwise will not. The measurement result for the all-zero state is always integer 0, and the state will not be changed. In the testing view, $U_{eo}^{-1}$ is the inverse variant of $U_{eo}$, which can be easily implemented as long as $U_{eo}$ can be easily implemented. The most significant advantage of this method is that only one run can get correct detection results with high probability. The disadvantage is that it is feasible only when a simple $U_{eo}$ of the expected state exists. If the code of $U_{eo}^{-1}$ is complex, then the TBD may be impractical. In this case, we need to use SBD or hybrid detection methods. Example~\ref{example:output} shows how to construct $U_{eo}^{-1}$ for the QFT program under classical input states.

\vspace{2mm}
\begin{example}\label{example:output} Output detection for QFT program.

The program specification of QFT has been shown in formula (\ref{EQUqftform}), and it can be rewritten in bitwise form:

\begin{equation} \label{EQUqftqubitform}
	\begin{array}{c}
		\left|j_1\right> \rightarrow \frac{1}{\sqrt 2}(\left|0\right>+e^{2\pi i0.j_n}\left|1\right>) \\
		\cdots \\
		\left|j_n\right> \rightarrow \frac{1}{\sqrt 2} (\left|0\right>+e^{2\pi i0.j_1j_2 \ldots j_n}\left|1\right>)
	\end{array}
\end{equation}

Given classical state input $\left|j\right> = \left|j_1\right>\cdots\left|j_n\right>$, the corresponding output state is the product state of $n$ single qubit states, and each single qubit state is in form $\frac{1}{\sqrt 2}(\left|0\right>+e^{i\theta_k}\left|1\right>)$, where $\theta_k = 0.j_{n-k+1} \ldots j_{n-1}j_n$. This single qubit state can be generated by applying gate $U_k$ on $\left|0\right>$, where

\begin{equation} \label{EQUgenqftqubit}
	U_k= R_1(\theta_k)H = \frac{1}{\sqrt 2} \left[
	\begin{array}{cc}
		1 & 1 \\
		e^{i\theta_k} & -e^{i\theta_k}
	\end{array}\right]
\end{equation}

To check the output, we can apply $U_k^{-1}=H R_1(-\theta_k)$ on $k$-th qubit. If the output is correct, all qubits after applying $U_k^{-1}$ will become $\left|0\right>$, then measuring the qubits will always get integer 0. Otherwise, non-zero results are possible to occur.

\end{example}
\vspace{2mm}

In fact, checking only the output states under classical state input cannot completely ensure the correctness of a quantum program. However, the output states under general input states may be complicated and hard to identify. Fortunately, if the target program is a unitary transform, the unitarity ensures that the correctness of the classical state input can deduce the correctness of the general input. So a feasible way to test a target unitary program \texttt{P} is:

\begin{itemize}
\item[1.] Check the output states of \texttt{P} under classical input states;
\item[2.] Perform additional unitarity checking for \texttt{P}.
\end{itemize}

Unfortunately, as we have investigated, hardly any work considers unitarity checking for quantum processes. It may be an important issue valuable to address.

\subsection{Testing Strategies for Variant Subroutines}
\label{subsec:testvariants}

In Section~\ref{subsec:Subroutines}, we discussed three variants of an original program. Essentially, they are different from the original program. Some programming languages support the management of these variants, such as automatically generating inverse or controlled variants. However, this automation is not always available, so testing for them is still necessary.

Fortunately, to test three variants of an origin subroutine, we can use the relation between the variants and the original one to avoid redesigning test cases, and the test can be finished in unified processes. Suppose we have finished testing the original subroutine \texttt{P}, we denote inverse, controlled, power variants of \texttt{P} as \texttt{InvP}, \texttt{CtrlP}, and \texttt{PowP}, respectively.

For \texttt{InvP}, there is an obvious relation:

\begin{equation}\label{equPInvPEqI}
\texttt{P}\circ\texttt{InvP}=I
\end{equation}

\noindent where `$\circ$' denotes the sequential execution (from right to left) of two subroutines, \texttt{InvP} and \texttt{P}, and $I$ is the identity operation. Note that no matter what \texttt{P} is, the relation (\ref{equPInvPEqI}) is always held, so we obtain a unified test process for any \texttt{InvP}, that is, to check whether the sequential execution of \texttt{InvP} and \texttt{P} is equivalent to identity $I$.

The input of \texttt{PowP} contains two parts: the power $k$ (a classical integer) and target qubits \textbf{qs}. Given $k$, the effect on \textbf{qs} is equivalent to apply \texttt{P} (if $k>0$) or \texttt{InvP} (if $k<0$) for $|k|$ times. Also, the equivalence check can be converted into an identity check with the following relations:

\begin{equation}\label{equPowP}
	\begin{array}{cc}
		\texttt{InvP}^k \circ \texttt{PowP}(k) = I,& k>0\\

		\texttt{P}^{|k|} \circ \texttt{PowP}(k) = I,& k<0
	\end{array}
\end{equation}

The input of \texttt{CtrlP} contains two parts: control qubits \textbf{qctrl} and target qubits \textbf{qtar}. If \textbf{qctrl} is in an \textit{all-one state} $\left|1\ldots1\right>$, then the effect on \textbf{qtar} is \texttt{P}. If \textbf{qctrl} is in a state which is orthogonal to $\left|1\ldots1\right>$, then the effect is identity $I$:

\begin{equation}\label{equCtrlP}
	\begin{array}{c}
		\texttt{InvP} \circ \texttt{CtrlP}(\textbf{qs} = \left|1\ldots1\right>) = I\\

		\texttt{CtrlP}(\textbf{qs} \bot \left|1\ldots1\right>) = I
	\end{array}
\end{equation}


We can see that the testing of variants subroutines can be converted into the checking of identity, which can be finished in unified processes. Here, $I$ has the program specification with the following form:

\begin{equation}\label{equIps}
I: \left|x\right> \rightarrow \left|x\right>
\end{equation}

\noindent
for all input states $\left|x\right>$. A feasible scheme is randomly selecting several input states and checking whether the program changes the input states.

\subsection{Integration Testing}
\label{subsec:Integration}


Similar to classical programs, bugs can also occur during the combination of subroutines, so integration testing should also be run for quantum programs. In this work, we use Q-UML~\cite{Perez-Delgado2020quantum} as guidance for our integration testing. The beginning of integration testing is to determine the order of integration. The strategies for classical integration testing can be migrated into quantum integration testing as well, such as topological sorting on the dependency graph to determine the integration order.

During the integration, if an upper subroutine needs to use a lower subroutine that is unfinished, we can use a \textit{test double} to replace it. This technique is suitable for testing quantum programs with oracles. Example~\ref{example:oracle} gives a strategy to test a program for searching quantum memory using the Grover algorithm. Test double can also be used to reduce costs and speed up the testing task. For example, if the testing task is run on a simulator, quantum arithmetic operations (such as adder or multiplier on qubits) are slow and consume too much memory. We can replace them with equivalent classical-quantum-mixed implementation to speed up the test process on the simulator.

\begin{figure}
	\centering
	\subfigure[Original Program] {\includegraphics[scale=0.6]{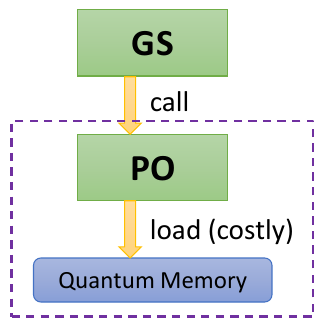}}\qquad
	\subfigure[Use test double $U_f$ to replace \texttt{PO} which loads quantum memory.] {\includegraphics[scale=0.6]{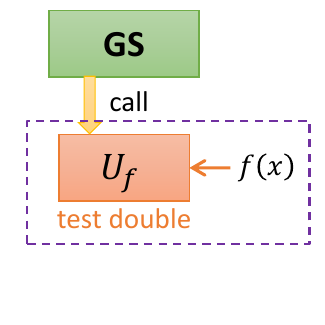}}
	\caption{The usage of test doubles to reduce the cost of testing \texttt{GS}.}
	\label{FIGtestdouble}
\end{figure}

\vspace{2mm}
\begin{example}\label{example:oracle} A quantum program that searches quantum memory by the Grover algorithm.
\noindent 

It has two subroutines: (1) general Grover Search subroutine \texttt{GS}, accepting an oracle; (2) subroutine \texttt{PO} to implement the oracle of \texttt{GS}, which loads the quantum memory, compares the contents, and flips the phase of corresponding amplitudes. In the whole program, \texttt{GS} calls \texttt{PO}, as Figure~\ref{FIGtestdouble}(a) shows.

We can test \texttt{PO} first and then combine it into \texttt{GS}. However, considering the effectiveness of locating bugs and the running cost (loading a quantum memory may be a costly operation), it is better to have individual testing for \texttt{GS}. In this case, the oracle is unfinished, and we should use test doubles to replace it. The test double can be implemented by some fixed unitary operations $U_{f}:\left|x\right> \rightarrow (-1)^{f(x)}\left|x\right>$, where the indicator function $f$ is chosen and controlled by testers, as Figure~\ref{FIGtestdouble}(b). Compared with the original \texttt{PO}, $U_{f}$ avoids loading quantum memory. Using $U_f$, \texttt{GS} can be tested individually at a lower cost.
\end{example}

\section{Tool Support}
\label{sec:QSharpTester}



It is necessary to build some tools which support finishing a test plan. This section first discusses the essential functions which should be provided by a testing tool and then introduces our testing tool \textbf{QSharpTester} for Q\# programs.

\subsection{Requirements of a Testing Tool}


According to the properties of quantum programs (Section~\ref{sec:basic-concepts}) and our testing methods (Section~\ref{sec:testing-framework}), we put forward four essential functions which should be provided by a testing tool for multi-subroutine quantum programs:

\begin{itemize} 
	\item[1.] Support of repeated execution of target programs;
	\item[2.] Support of data processing for quantum information;
	\item[3.] Support of composition of quantum subroutines;
	\item[4.] Support of results analysis.
\end{itemize}

Function 1 is based on the fact that SBD methods (Section~\ref{subsec:GenerationDetection}) are important in detecting output and require the repeated execution of target programs, while Function 3 is for multi-subroutine quantum programs. These four functions are the minimum requirements for a testing tool, and a practical testing tool should provide more additional functions.

\subsection{QSharpTester: A Testing Tool for Multi-Subroutine Q\# Programs}

Q\# provides the package \texttt{Microsoft.Quantum.XUnit} for users to perform unit testing. It is only a basic framework supporting test scripts and showing results. Therefore, it is necessary to build a more powerful testing tool. We implemented QSharpTester,
a testing tool for Q\# programs, which includes the following libraries:

\noindent 
\begin{itemize} 
    \item \textbf{Preparation}: It provides methods for preparing common and important input quantum states.
    \item \textbf{Detection}: It provides the mechanism for running the target program repeatedly and some common methods to detect output testing.
    \item \textbf{Assertion}: It provides some methods for users to write test scripts.
    \item \textbf{Testing}: It provides some preset procedures for performing specific tests, including the test of inverse or controlled variants.
    \item \textbf{Display}: It provides some methods to display the testing results.
\end{itemize}

A complete testing task for Q\# programs can be finished using QSharpTester and \texttt{Microsoft.Quantum.XUnit}. In the following section, we will evaluate the effectiveness of our testing methods using QSharpTester.

\section{Evaluation}
\label{sec:evaluation}

\begin{table*}
\footnotesize{
	\centering
	\caption{Benchmark programs}
	\label{table:BenchmarkProg}
	\begin{tabular}{cc|cc|cccc}
		\toprule
		\multirow{2}{*}{\makecell[c]{\textbf{Related}\\ \textbf{RQs}}} & \multirow{2}{*}{\textbf{Program}} & \multirow{2}{*}{\textbf{IO mark}} & \multirow{2}{*}{\textbf{Program Specification}} & \multicolumn{4}{c}{\textbf{Number of mutations}} \\
		&&&& GM & SM & CM & MM\\
		\toprule
		
		\multirow{6}{*}{RQ1, RQ2}& \texttt{Reverse} & ($n$, \textbf{qs}) $\rightarrow$ (\textbf{qs'})  & $\left|j_1\right>\dots\left|j_n\right> \rightarrow \left|j_n\right>\dots\left|j_1\right>$ & 16 & 0 & 12 & 10 \\
		\cmidrule{2-8}
		& \texttt{MultiSWAP} & ($n$, \textbf{qs1}, \textbf{qs2}) $\rightarrow$ (\textbf{qs1'}, \textbf{qs2'}) & $\left|\phi\right>\left|\psi\right>\rightarrow\left|\psi\right>\left|\phi\right>$ & 13 & 0 & 8 & 10\\
		
		\cmidrule{2-8}
		& \texttt{QFT} & ($n$, \textbf{qs}) $\rightarrow$ (\textbf{qs'})  & $	\left|j\right> \rightarrow \frac{1}{\sqrt{2^n}}\sum_{k=0}^{2^n-1}e^{2\pi i j k / 2^n}\left|k\right>$ & 20 & 16 & 18 & 12 \\
		\cmidrule{2-8}
		& \texttt{invQFT} & ($n$, \textbf{qs}) $\rightarrow$ (\textbf{qs'})  & * & 18 & 16 & 16 & 12 \\
		
		\midrule

		\multirow{5}{*}{RQ2}& \texttt{Purity} & ($n$, $t$, \underline{GenRho}) $\rightarrow$ (isPure)  & return TRUE if $\rho$ is pure state. & 0 & 6 & 7 & 4 \\
		\cmidrule{2-8}
		
		& \texttt{PhaseFlip} & ($n$, \textbf{qs}) $\rightarrow$ (\textbf{qs'})  & $	\left|x\right> \rightarrow \left\{
			\begin{array} {cc}
				-\left|x\right> & x>0\\
				\left|x\right> & x=0
			\end{array}
			\right.$ & 7 & 0 & 3 & 4\\
		\cmidrule{2-8}
		& \texttt{Grover} & ($n$, \underline{OracleK}) $\rightarrow$ (\textbf{qs'}) & $\left|0\right> \rightarrow \  \approx \left|K\right>$ & 3 & 8 & 2 & 3 \\

		\bottomrule
	\end{tabular}

}
\end{table*}

We evaluate our testing methods with the following research questions:

\begin{itemize} 
    \item\textbf{RQ1:} Is covering both classical and superposition states necessary?
    
    \item\textbf{RQ2:} How do our testing methods perform on the benchmark programs?
    
    \item\textbf{RQ3:} Compared with existing testing methods, what are the advantages of our methods?
\end{itemize}

\subsection{Benchmark Programs}
\label{subsec:Benchmarks}

We use the following seven original Q\# programs as a benchmark to evaluate the effectiveness and necessity of our testing methods.

\begin{enumerate}[label={(\arabic*)}]
    \item \texttt{Reverse}: It reverses the order of qubits (see lines 10-17 in Listing~\ref{LISTqft}).
    
    \item \texttt{MultiSWAP}: It swaps quantum variables.
    
    \item \texttt{QFT}: It performs the Quantum Fourier transform (see lines 19-33 in Listing~\ref{LISTqft}). 
    
    \item \texttt{invQFT}: It performs the inverse Quantum Fourier transform.
    
    \item \texttt{Purity}: Given a subroutine that generates a given state $\rho$, it returns whether $\rho$ is a pure state by several repetitions.
    
    \item \texttt{PhaseShift}: A subroutine in \texttt{Grover}. 
    
    \item \texttt{Grover}: Given an oracle, it performs the Grover Search algorithm.
\end{enumerate}

For each original program, we construct several faulty programs by mutating the program using the following four types of mutations:

\begin{itemize} 
    \item \textbf{Gate Mutation (GM)} has four types of operations: adding, removing, modifying, and reordering the basic quantum gates in the program.
    
    \item \textbf{Subroutine Mutation (SM)} has four types of operations as well, including adding, removing, and modifying the call statements in the program, which calls other subroutines and also especially replacing the subroutine with an error version.
    
    \item \textbf{Classical Mutation (CM)} contains the modification of the assignment of classical variables and control flows.
    
    \item \textbf{Measurement Mutation (MM)} has three types of operations: adding, removing, and modifying the measurement statements in the program.
\end{itemize}

Note that compared to previous research on testing quantum programs~\cite{ali2021assessing,wang2021generating,wang2021application,fortunato2022qmutpy} and quantum platforms~\cite{wang2021qdiff}, which mainly considers the gate mutations (GMs) as well as measurement mutations (MMs), our research uses more types of mutations that are commonly used in practical development to evaluate our testing method.

Table \ref{table:BenchmarkProg} shows our benchmark programs in detail, including their IO marks, program specifications, and the number of each type of mutation. For a benchmark program, the number of a type of mutation may be 0. For example, there is no SM for a program that does not call other subroutines.

\subsection{Experiment Design and Settings}
\label{subsec:ExpSet}

In our experiment, as Table~\ref{table:BenchmarkProg} shows, we use the first four programs (\texttt{Reverse}, \texttt{MultiSWAP}, \texttt{QFT}, and \texttt{invQFT}) to answer RQ1 and all seven programs to answer RQ2. The reason is that the first four programs are suitable for using the CSP criterion.

We use the CSP criterion to partition the quantum input states of \texttt{Reverse}, \texttt{MultiSWAP}, and \texttt{QFT}. The scale of \texttt{Reverse} is divided into $n=1$ and $n>1$, and we choose $n=6$ for the $n>1$ class. As shown in Example~\ref{example:qftpart}, the scale of \texttt{QFT} and \texttt{invQFT} should be divided into $n=1$, $n=2$, and $n\ge 3$, and we also choose $n=6$ for the $n\ge 3$ class. We use ACoC to combine variable $n$ and \textbf{qs} in \texttt{Reverse} and \texttt{QFT}, which means that for each condition of $n$, we test both classical and superposition inputs. \texttt{MultiSWAP} has three input variables: $n$, \textbf{qs1}, and \textbf{qs2}. As our goal is to answer RQ1, here we only consider the inputs that \textbf{qs1} and \textbf{qs2} are both classical or superposition states.

For \texttt{invQFT}, we assume that we have finished the testing of \texttt{QFT} and then use relation (\ref{equPInvPEqI}) to covert the testing task into identity check: $\texttt{QFT}\circ\texttt{invQFT}=I$.

We generate $2n^2$ test cases for each equivalent class (i.e., 2 cases for $n=1$, 8 cases for $n=2$, and 72 cases for $n=6$), and use them on each faulty program. Here, we chose the number $2n^2$ based on our practical experiences considering a balance between accuracy and running time. We generate each test case randomly with the support of QSharpTester. Concretely, for the classical input state class (CI), we use input state $\left|x\right>$, where $x$ is a random $n$-bits binary string. For superposition input state class (SI), we use input state $\frac{1}{\sqrt 2}(\left|x\right>+\left|y\right>)$, where $x$ and $y$ are two $n$-bits binary strings. The experiment settings for these four programs are shown in Table~\ref{table:Necessity}.

For \texttt{Purity}, the input parameter \underline{GenRho} is a subroutine to generate target $\rho$, and $t$ is the number of repetitions. \texttt{Purity} contains subroutine \texttt{MultiSWAP}, which estimates the purity of the target state $\rho$ by swap test~\cite{buhrman2001quantum,ekert2002direct}. If there is a round obtaining result ``mixed," \texttt{Purity} will return FALSE; otherwise, TRUE. We fix $t=10$. Note that \texttt{Purity} will run under mixed $\rho$ so that we can partition \underline{GenRho} by the CSMP criterion, although it is not a quantum variable.

For \texttt{PhaseFlip}, since \textit{global phases} cannot be found by measurement~\cite{nielsen2002quantum}, we should have a fine-grained partition rather than a coarse-grained CSP or CSMP. According to program specification, it is feasible to partition \textbf{qs} into 4 types: (1) $\left|0\right>$\footnote{Here $\left|0\right>$ represents a classical state with integer 0, rather than a single qubit.}; (2) $\left|x_1\right>, x_1>0$; (3) $\frac{1}{\sqrt{2}}(\left|0\right>+\left|x_1\right>), x_1>0$; (4) $\frac{1}{\sqrt{2}}(\left|x_1\right>+\left|x_2\right>), x_1,x_2>0, x_1 \neq x_2$. The output of \texttt{PhaseFilp} under input types (1)(2)(4) can not be distinguished, whereas (3) can.

For \texttt{Grover}, as Example~\ref{example:oracle} shows, we use some test doubles to fill the subroutine-type input \underline{OracleK}.

We use Q\# language and its simulator to conduct our experiments since (1) our methods and benchmark programs are variable-scale and classical-quantum-mixed, and (2) Q\# is a high-level quantum programming language that supports this pattern of programs. We run experiments on a personal computer with an Intel Core i5-10210U CPU and 16 GB RAM.

\subsection{RQ1: The Necessity of Covering Both Classical and Superposition States}\label{subsec:necessity}

Table~\ref{table:Necessity} shows the number of trigger test cases. Here ``trigger" means the faulty program is reported as a ``fail" under the test case.

We use SBD methods only for the superposition input states of \texttt{QFT}, and QRA methods for other cases because no QRA method is easy to implement for the output of \texttt{QFT} under superposition input. The number of repetitions of SBD is set to 200. The number we choose here is also based on practical experiences considering the balance between accuracy and running time.

Concretely, \texttt{Reverse} exchange the big endian and little endian of a qubit array. QSharpTester provides some methods that can generate or uncompute given states on a qubit array with a given endian mode. If the \texttt{Reverse} is correct, the output state under the input state, which is generated by little endian mode from the all-zero state, can be uncomputed into the all-zero state by big-endian mode. \texttt{MultiSWAP} exchanges two-qubit arrays with the same length. Suppose $G_1$ and $G_2$ generate the input states of \textbf{qs1} and \textbf{qs2}; the output states can be uncomputed into all-zero states by $G_2^{-1}$ and $G_1^{-1}$, respectively. How to check the output state for \texttt{QFT} under classical input, denoted as  $\left|\psi_{j}\right> = \texttt{QFT}(\left|j\right>) = \frac{1}{\sqrt{2^{n}}}\sum_{k=0}^{2^{n}-1}e^{2\pi i j k / 2^{n}}\left|k\right>$, has been shown in Example~\ref{example:output}. What remains is how to detect output under a two-value-superposition input state, which has the form $\frac{1}{\sqrt 2}(\left|\psi_{a}\right>+\left|\psi_{b}\right>)$, where $a$ and $b$ are two binary strings. There is no intuitive way to generate this state, but it has a useful property. Suppose for any $j$ we have $U_{j}$ such that $U_j\left|0\ldots0\right>=\left|\psi_j\right>$. Apply $U_a^{-1}$ on the target state: $U_a^{-1}[\frac{1}{\sqrt 2}(\left|\psi_{a}\right>+\left|\psi_{b}\right>)] = \frac{1}{\sqrt 2}(\left|0\ldots0\right>+U_a^{-1}\left|\psi_{b}\right>)$. Then about half of the measurement results will be 0. $U_b^{-1}$ is similar. We can use statistical methods to check whether the 0 results occur in about half. Specifically, we give a tolerance of 0.1, which means the test will pass if the frequency of 0 results is between 0.4 and 0.6.

From the experiment, we can see that due to the repetitions, the testing tasks running by SBD are slower than that of QRA, but the trigger rate by SBD is higher than that of QRA. The reason is that QRA is not a deterministic algorithm. QRA only ensures it will report a ``pass" when the output is correct. It still has some probability of reporting a “pass” (false negative) for the wrong output. A better method is to repeat the QRA process several times, just like other random algorithms. If there is at least one case fail, report ``fail." As long as the probability of returning correct results is not exponentially small, a reliable result can be obtained after acceptable repetitions~\cite{arora2009computational}.

In the $n=1$ class, the trigger rate of each program is always low. The reason is that $n=1$ means not executing loops, so the mutations inside of loops will not be triggered. According to the result of $n=1$ test cases, we know whether the bugs are inside or outside the loops.

Interestingly, the trigger rate of MMs by classical input states is significantly lower than that of MMs by superposition input states. Especially for \texttt{Reverse} and \texttt{MultiSWAP}, classical input states cannot trigger any MMs. We have further studied the phenomenon and discovered: (1) If the program is classical-like, which means it only contains ``essentially classical" gates (such as X, CNOT, and SWAP), classical input states cannot trigger any MMs. It is because under classical input states, the computation process is essentially classical, and measurement will not change the contents of variables. (2) If the MM occurs at the beginning of a program, classical input states cannot trigger it. The reason is that measurement will not change the given input. (3) Some ``phase-only" mutations (such as adding an S or Z gate) cannot be triggered by the classical input states.

Similarly, we also found that some mutations that can be triggered by classical input states can not be triggered by superposition input states. This fact implies that superposition is one of the essential differences between classical and quantum computing. Thus, covering both classical and superposition inputs is necessary for testing quantum programs.

\begin{table*}
\scriptsize{
	\centering
	\caption{RQ1: Necessity evaluation for covering both classical(CI) and superposition(SI) input states}
	\label{table:Necessity}
	\begin{tabular}{c|cc|c|cc|ccccc|ccccc|cc}
		\toprule
		\multirow{2}{*}{\textbf{Program}} & \multirow{2}{*}{\makecell[c]{\textbf{Criteria/} \\ \textbf{Methods}}} & \multirow{2}{*}{\makecell[c]{\textbf{Detection} \\ \textbf{Method}}} & \multirow{2}{*}{$n$} & \multicolumn{2}{c}{\#(Test Cases)} & \multicolumn{4}{c}{\#(Trigger by CI)} & \multirow{2}{*}{\makecell{trigger\\rate}} & \multicolumn{4}{c}{\#(Trigger by SI)}& \multirow{2}{*}{\makecell{trigger\\rate}} & \multicolumn{2}{c}{\textbf{Run Time}} \\
		&&&&CI&SI&GM&SM&CM&MM& &GM&SM&CM&MM& &CI&SI\\
		\toprule
		
		\multirow{2}{*}{\texttt{Reverse}} & \multirow{2}{*}{CSP+ACoC} & \multirow{2}{*}{QRA} & 1 & 76 & 76 & 9 & - & 6 & 0 & 19.7\% & 9 & - & 6 & 5 & 26.3\% & \multirow{2}{*}{6.1s} & \multirow{2}{*}{7.5s}\\
		&&& 6 & 2736 & 2736 & 831 & - & 705 & 0 & 56.1\% & 963 & - & 759 & 288 & 73.5\% &&\\
		
		\midrule
		
		\multirow{2}{*}{\texttt{\tiny{MultiSWAP}}} & \multirow{2}{*}{CSP+ECC} & \multirow{2}{*}{QRA} & 1 & 62 & 62 & 18 & - & 7 & 0 & 40.3\% & 8 & - & 6 & 10 & 38.7\% & \multirow{2}{*}{13.0s} & \multirow{2}{*}{16.2s} \\
		&&& 6 & 2232 & 2232 & 844 & - & 535 & 0 & 61.8\% & 883 & - & 549 & 330 & 78.9\% && \\
		
		\midrule
		
		\multirow{3}{*}{\texttt{QFT}} & \multirow{3}{*}{CSP+ACoC} & \multirow{3}{*}{\makecell[c]{QRA for CI\\SBD for SI}} & 1 & 132 & 132 & 6 & 1 & 3 & 4 & 10.6\% & 12 & 0 & 4 & 4 & 15.2\% & \multirow{3}{*}{18.3s} & \multirow{3}{*}{\makecell[c]{155 \\ min}} \\
		&&& 2 & 528 & 528 & 86 & 63 & 67 & 44 & 49.2\% & 109 & 96 & 92 & 84 & 72.2\% && \\
		&&& 6 & 4752 & 4752 & 1186 & 969 & 1135 & 493 & 79.6\% & 1391 & 1077 & 1290 & 830 & 96.5\% && \\
		
		\midrule
		
		\multirow{3}{*}{\texttt{invQFT}} & \multirow{3}{*}{\makecell{Use\\ relation~(\ref{equPInvPEqI}) }} & \multirow{3}{*}{\makecell[c]{\tiny{Check whether}\\ \tiny{$\texttt{QFT} \circ \texttt{invQFT}$ } \\ $=I$ }} & 1 & 124 & 124 & 3 & 0 & 4 & 3 & 8.1\% & 7 & 0 & 4 & 3 & 11.3\% & \multirow{3}{*}{24.2s} & \multirow{3}{*}{27.9s} \\
		&&& 2 & 496 & 496 & 57 & 52 & 40 & 40 & 38.1\% & 78 & 62 & 52 & 39 & 46.6\% && \\
		&&& 6 & 4464 & 4464 & 943 & 1062 & 1133 & 595 & 83.6\% & 1086 & 954 & 1233 & 718 & 89.4\% && \\
		
		\bottomrule
	\end{tabular}
}
\end{table*}

\subsection{RQ2: Performance}
\label{subsec:discover}

To show the performance of our testing methods, we use all seven benchmark programs in Table~\ref{table:BenchmarkProg}. Like Section~\ref{subsec:necessity}, we use random testing to generate $2n^2$ test cases for each equivalence class. The mutation program is killed if one test case of all equivalence classes fails. 
Table~\ref{table:AllResults} gives the number of unkilled mutations for each benchmark program. Almost all mutations are killed, which means our testing methods perform well.

The last column of Table~\ref{table:AllResults} shows the run time of testing each program. We can see that the testing tasks can be finished on a personal computer and Q\# simulator at an acceptable time. The time is related to the number of repetitions of the measurement in each test case.

We have further studied the unkilled mutations. For the unique unkilled SM of \texttt{QFT}, the concrete mutation is:

\vspace{1mm}
\footnotesize{
\lstinline|-    CRk(qs[j], k, qs[i]);       //correct|

\lstinline|+    CRk(qs[i], k, qs[j]);       //mutation|
}
\vspace{2mm}

\normalsize
\noindent The unique unkilled CM of \texttt{PhaseFlip} is:

\vspace{1mm}
\footnotesize{
\lstinline|-    Controlled Z(qs[0 .. N - 2], qs[N - 1]);|

\lstinline|+    Controlled Z(qs[1 .. N - 1], qs[0]);|
}
\vspace{2mm}

\normalsize
We can prove that the two mutations are equivalent to the correct program. Actually, they are \textit{controlled Z-axis rotation} operations. Here we consider the case of one control qubit and one target qubit. The general matrix is $diag\{1,1,1,e^{i\theta}\}$, and the status of the control qubit is equivalent to that of the target qubit. For example, \texttt{CRk} is a case with $\theta=\frac{1}{2^k}$, and the first mutation is derived by exchanging the control and target qubits. The equivalence of the second mutation is for the same reason. Note that \texttt{qs[1]$\cdots$qs[N-2]} are always controlled qubits before and after the mutation. The difference is just for qs[0] and qs[N-1]. The mutation is just exchanging the control/target status of qs[0] and qs[N-1], so they are equivalent.

The unique unkilled MM of \texttt{Purity} is adding measurement after resetting all qubits, so the redundant measurement will not change the state of qubits.

\begin{table}
\scriptsize{
	\centering
	\caption{RQ2: Run time and the number of unkilled benchmark programs}
	\label{table:AllResults}
	\begin{tabular}{c|c|cccc|c|c}
		\toprule
		\multirow{2}{*}{\textbf{Program}} & \multirow{2}{*}{\makecell[c]{\#(Total \\ Mutations)} } & \multicolumn{5}{c}{\textbf{\#(unkilled mutations)}} & \multirow{2}{*}{\makecell[c]{Run \\ Time}} \\
		&&GM&SM&CM&MM&sum&\\
		\toprule
		\texttt{Reverse} & 38 & 0 & 0 & 0 & 0 & 0 &13.6s\\
		\texttt{MultiSWAP} & 31 & 0 & 0 & 0 & 0 & 0 & 29.2s\\
		\texttt{QFT} & 66 & 0 & \textbf{1} & 0 & 0 & 1 & 156min\\
		\texttt{invQFT} & 62 & 0 & 0 & 0 & 0 & 0 & 52.1s\\
		\texttt{Purity} & 17 & 0 & 0 & 0 & \textbf{1} & 1 & 11.2min\\
		\texttt{PhaseFlip} & 14 & 0 & 0 & \textbf{1} & 0 & 1 & 4.5s\\
		\texttt{Grover} & 16 & 0 & 0 & 0 & 0 & 0 & 2.2min\\
		\bottomrule
	\end{tabular}
}
\end{table}

\subsection{RQ3: Compared with Existing Methods}
\label{subsec:advantages}

Our research aims to develop a novel framework for supporting the test of multi-subroutine quantum programs rather than a concrete method for some types of programs. It means that our work and existing methods are not in the same hierarchy, so we cannot compare their execution efficiency. However, comparing them from supporting users to finishing testing tasks is still valuable. 

\vspace{2mm}
\subsubsection{Support Wider Program Types}
Compared with existing methods, the most significant advantage of our work is to support broader types of quantum programs, which contain quantum programs with complex subroutine structures, variable scales, and complex parameters. Existing methods~\cite{ali2021assessing, wang2021generating, wang2021application} focus mainly on the quantum circuit level, and their definitions of input and output limit the scope of the application.

As Tables~\ref{table:BenchmarkProg} and~\ref{table:AllResults} show, our methods can deal with not only \textit{usual} quantum programs such as \texttt{Reverse} and \texttt{QFT}, which can also be dealt with by existing methods, but also programs with complex parameters, such as \texttt{MultiSWAP} (a program with multiple quantum input variables), \texttt{Grover} (a program with oracle), and \texttt{Purity} (a program has even no quantum input variables).

\vspace{2mm}
\subsubsection{Quantum Natures}
Unlike existing methods that migrate classical testing methods into the quantum domain simply, our work considers more the nature of a quantum program. We propose CSP and CSMP criteria that partition quantum variables by quantum nature rather than considering only the classical values. As a result, our methods can find bugs related to the nature of a quantum program, which existing methods cannot. For example, existing methods cannot test \texttt{PhaseFlip}.

\vspace{2mm}
\subsubsection{Scalability and Compatibility}
As shown in Table~\ref{table:Necessity}, existing classical partition and combination coverage criteria (such as boundary value partition and ACoC combination criterion) and both QRA and SBD detection methods can be used in our framework. In addition, our framework can be compatible with existing methods. For example, we can use search-based methods~\cite{wang2021generating} to generate test cases rather than manual construction in some concrete tasks. The evaluation demonstrates the scalability and compatibility of our framework.

\section{Related Work}
\label{sec:related-work}


Quantum software testing is an emerging research field still in the preliminary stage~\cite{zhao2020quantum}. Miranskyy and Zhang~\cite{miranskyy2019testing} discussed the importance and challenges of testing quantum programs. Miranskyy {\it et al.}~\cite{miranskyy2021testing} and Garc{\'\i}a de la Barrera {\it et al.}~\cite{garcia2021quantum} presented the research progress in quantum software debugging and testing.

A typical thread of research in this area is to apply well-established testing methods for classical programs to the domain of quantum computing. Wang {\it al et.}~\cite{wang2018quanfuzz} proposed QuanFuzz, a fuzzy testing method for generating test cases for quantum programs. Honarvar {\it et al.}~\cite{honarvar2020property} proposed a property-based testing method to test the Q\# program and developed a tool called QSharpChecker. Li {\it et al.}~\cite{li2020projection} proposed Proq, a projection-based runtime assertion tool for testing and debugging quantum programs. Proq uses projection to represent assertions, which, compared to classical representation, has more expressive power. Ali {\it et al.}~\cite{ali2021assessing} defined the input-output coverage criteria for quantum program testing and used mutation analysis to evaluate the effectiveness of the criteria. 
Wang {\it et al.}~\cite{wang2021generating} proposed QuSBT, a search-based test case generation algorithm to generate test cases containing more failure cases. They also proposed a combinatorial testing approach called QuCAT~\cite{wang2021application} to test quantum programs through testing cases of input variables of the programs. The basic idea of the approach is to automatically generate test suites of a specific size based on the available test budget to maximize the number of failed test cases in the test suite. Abreu {\it et al.}~\cite{abreu2022metamorphic} proposed a metamorphic testing method for oracle quantum programs. The basic idea is to define some metamorphic rules specific to quantum programs and use these rules to support the testing of quantum programs. Some researchers also apply mutation testing and analysis techniques to the domain of quantum computing to support testing quantum programs~\cite{fortunato2022qmutpy,mendiluze2021muskit}. Compared to this thread of research, which mainly applied specific testing techniques to quantum programs, our testing framework can support a complete and systematic test process for multi-subroutine quantum programs. 

Assertions can be used to check quantum states and help users find bugs in quantum circuits and programs. A straightforward idea to implement quantum assertions is to prepare many copies of quantum variables and repeat the measurement. Huang {\it et al.}~\cite{huang2019statistical} applied hypothesis testing to partially reconstruct the information of quantum registers according to the measurement results and implement assertion. However, the measurement-based method cannot implement runtime assertion.
Liu {\it et al.}~\cite{liu2020quantum} proposed that additional qubits can be introduced to capture the information of the qubits under assertion, so there is little interruption for the program execution during assertion. 
Li {\it et al.}~\cite{li2020projection} proposed a projection-based runtime assertion scheme, using the projection operator to represent the predicates in the assertion. Liu {\it et al.}~\cite{liu2020quantum} extended this idea to the assertion of mixed states and approximate assertion. In this paper, we follow these ideas to perform output detection of quantum programs. 

Building some benchmarks to assess the effectiveness of different testing methods is necessary. Campos and Souto~\cite{campos2021qbugs} proposed QBugs, a collection of reproducible bugs in quantum algorithms for supporting controlled experiments for quantum software debugging and testing. But it is not available to assess its details and usability. Zhao {\it et al.}~\cite{zhao2021bugs4q} proposed Bugs4Q, an open-source benchmark of 36 real, validated bugs in practical Qiskit programs, supplemented with the test cases for reproducing buggy behaviors for supporting quantum program testing.  
As these benchmarks become mature, we can consider using them to evaluate our testing method.

\section{Conclusion}
\label{sec:conclusion}
This paper has proposed a novel framework for testing quantum programs with multiple subroutines. Our framework contains IO analysis, equivalence class partition, coverage criteria, and test case generation. To support our framework, we also implemented QSharpTester, a testing tool for Q\# programs, and conducted experiments using Q\# programs to demonstrate the effectiveness of our testing framework. Our experiment used a set of Q\# benchmark programs, including seven original correct programs and their 244 faulty mutations with four mutation types. The experimental results show that our methods can find almost all faulty mutations and some equivalent mutations and can finish testing tasks in an acceptable time on a personal computer with a Q\# simulator. 
Moreover, although we present our testing framework in the context of Q\#~\cite{svore2018q}, the underlying ideas are applicable to other quantum programming languages such as Scaffold~\cite{abhari2012scaffold}, Silq~\cite{bichsel2020silq}, and Qiskit~\cite{gadi_aleksandrowicz_2019_2562111}.

In future work, we plan to investigate several critical issues further in testing quantum programs, including (1) giving a broader definition of quantum program specifications, (2) defining more novel coverage criteria for quantum variables
, and (3) further investigating metamorphic testing of quantum programs.



\bibliographystyle{acm}
\bibliography{ref}

\end{document}